\documentclass[usenatbib,letters]{mnras}
\usepackage{amsmath}
\usepackage{graphicx}
\usepackage{amssymb,txfonts}
\usepackage[T1]{fontenc}

\title[Outflowing lamp-post]{Lamp-post with an outflow and the hard state of Cyg X-1}

\author[Klepczarek et al.]
{{\L}ukasz Klepczarek$^1$, Andrzej Nied{\'z}wiecki$^1$ and Micha{\l} Szanecki$^1$\\
$^1$Faculty of Physics and Applied Informatics, {\L}{\'o}d{\'z} University, Pomorska 149/153, 90-236 {\L}{\'o}d{\'z}, Poland\\
}

\pagerange{\pageref{firstpage}--\pageref{lastpage}}
\pubyear{2022}

\begin{document}

\defcitealias{2019MNRAS.485.2942N}{N19}
\defcitealias{2015ApJ...808....9P}{P15}
\defcitealias{2017MNRAS.472.4220B}{B17}
\defcitealias{1999ApJ...510L.123B}{B99}

\maketitle

\label{firstpage}

\begin{abstract}
Relativistic reflection observed in the hard states of accreting black holes (BH) often shows a weak amplitude relative to the main Comptonization component, which may result from either a disc truncation or a non-isotropy of the X-ray source, e.g.\ due to a motion away from the reflector. We investigate here the latter case, assuming that the X-ray source is located on the symmetry axis of the Kerr BH. We discuss effects relevant to a proper computation of the reflected radiation and we implement them in the model for data analysis, {\tt reflkerrV}. We apply it to the simultaneous {\it Suzaku} and {\it NuSTAR} observation of Cyg X-1 in the hard state and we find a good fit for an untruncated disc irradiated by the source moving away from it at 0.36c. However, we find a slightly better solution in a geometry closely approximating the truncated disc irradiated by an inner hot flow. In this solution we either still need a subrelativistic outflow or the source opposite to the observer must  contribute to the directly observed radiation. We also discuss differences between the implementation of the outflow effect in {\tt reflkerrV} and in {\tt relxilllpCp}.
\end{abstract}


\section{Introduction}
\label{sect:intro}

Relativistic reflection spectroscopy is an important tool to study the morphology of the accreting matter in the vicinity of BH horizon \citep[e.g.][]{2021SSRv..217...65B}. A popular reflection model involves a compact source on the symmetry axis of the accretion system, so-called lamp post (LP). This geometry, first considered by \cite{1996MNRAS.282L..53M} for estimating the enhanced irradiation of the inner disc by gravitational focusing in the mathematically simplest case, has been then used to compute the spectrum of the reflected radiation in a number of works \citep[e.g.][]{2004MNRAS.349.1435M,2008MNRAS.386..759N,2013MNRAS.430.1694D}. An important motivation for the LP geometry comes from attempts of constraining the BH spins using the relativistic reflection spectroscopy \citep[e.g.][]{2021ARA&A..59..117R}. These estimations typically require very centrally-concentrated profiles, which can only be explained if the X-ray source is located very close to the BH horizon. While severe problems were pointed out for the LP geometry, 
especially for extreme (close to event horizon) locations of the X-ray source \citep{2016ApJ...821L...1N}, it has an important advantage of predicting the spectral distortions by relativistic effects of the Kerr metric within their physical limits. This is in contrast to phenomenological models assuming an arbitrary radial distribution of reflection, in which nonphysically steep emissivities are often estimated, leading to nonphysical conclusions \citep[see discussion in][]{2020A&A...641A..89S}.

A source on the symmetry axis can be associated with the region of jet formation \citep[e.g.][]{2004A&A...413..535G} and then it is likely that such a source moves outward along the symmetry axis. Such a motion was considered by \citet{2013MNRAS.430.1694D} and it is implemented in their {\tt relxilllpCp} model. The model has been applied e.g. by \citet{2021A&A...654A..89P} and \citet{2021NatCo..12.1025Y} to find velocity of the X-ray source. The other (public) relativistic reflection models, i.e.\ {\tt kyn} \citep{2004ApJS..153..205D},  {\tt reltrans}  \citep{2019MNRAS.488..324I} and {\tt reflkerr} \citep[][hereafter \citetalias{2019MNRAS.485.2942N}]{2019MNRAS.485.2942N}, and also the recent version 2.0 of {\tt relxilllpCp}, allow to scale the predicted strength of reflection to approximate the effect of non-isotropic X-ray emission. 
This approach, however, neglects other effects resulting from the non-isotropy, in particular, the change of the radial emissivity profile \citep[see e.g.\ fig.\ 9 in][]{2007ApJ...664...14F}. Then, spectra predicted by these models for the scaling parameter significantly different from unity are not self-consistent and the results of their application may be unphysical. In particular, a non-isotropy sufficiently large to yield a substantial reduction of reflection would also remove the steep part of the profile  close to the BH,  characteristic to the LP model \citep[see fig.\ 5 in][]{2016ApJ...821L...1N}, which effect is not accounted for in models with a free scaling of reflection.

 In this work we reconsider the effect of vertical motion of the X-ray source. We implement it in a new {\tt xspec} model, {\tt reflkerrV}. In Section \ref{sect:model} we describe this model, we compare it with {\tt relxilllpCp} and we discuss  differences between these models.

An important application of the model with an outflowing X-ray source concerns the hard states of BH binaries, in which the observed reflection component is often weaker by a factor of several than expected for an isotropic source above an untruncated disc, e.g.\ \citet{1997MNRAS.288..958G}, \citet{2001ApJ...547.1024D}, \citet{2015ApJ...813...84G}, \citet[][hereafter \citetalias{2015ApJ...808....9P}]{2015ApJ...808....9P}, \citet[][hereafter \citetalias{2017MNRAS.472.4220B}]{2017MNRAS.472.4220B}. This can be explained if reflection arises from either a truncated disc irradiated by a hot inner flow \citep{1976ApJ...204..187S, 1998ApJ...505..854E}, or a disc irradiated by a source moving away from it with mildly relativistic velocity \citep[][hereafter \citetalias{1999ApJ...510L.123B}]{1999ApJ...510L.123B}. In Section \ref{sect:cygx1} we apply our reflection model to the hard state observation of Cyg X-1 by {\it Suzaku} and {\it NuSTAR} to find if the current quality of X-ray spectral data allows to discriminate between these two solutions.

\section{LP with a vertical motion}
\label{sect:model}

We consider an X-ray source on the symmetry axis of a Kerr BH with the spin $a$. The BH is surrounded by a Keplerian disc which may be either truncated at the inner radius $r_{\rm in}$ or extend to the innermost stable circular orbit (ISCO). We assume that the emission region is static, however, electrons in it undergo a bulk motion in the direction perpendicular to the disc with velocity $v$ as measured in the locally non-rotating frame \citep[LNRF,][]{1972ApJ...178..347B}. In this assumption we follow the physical model of \citetalias{1999ApJ...510L.123B} with an X-ray source dominated by $e^\pm$ pairs, which are ejected away by the radiation pressure, immediately cool down and are replaced by newly created pairs.

 The model \texttt{reflkerrV} developed here is included in the \texttt{reflkerr} family and we strictly follow the theoretical framework described in \citetalias{2019MNRAS.485.2942N}, i.e.\ we construct the source-to-disc ($\mathcal{T}_{\rm sd}$) and source-to-observer ($\mathcal{T}_{\rm so}$) transfer functions  by tabulating a large number of photon trajectories. We assume that the intrinsic X-ray emission is isotropic and we generate photons with an isotropic distribution of initial directions in the frame co-moving with electrons. We then make the Lorentz transformation to the LNRF frame, we find the constants of motion and we compute the photon trajectory using the method of \citet{2008MNRAS.386..759N}. The fluxes of photons reaching the disc, tabulated in $\mathcal{T}_{\rm sd}$, and directly observed, tabulated in $\mathcal{T}_{\rm so}$, are affected by the relativistic aberration and Doppler effects, however, there is no retardation effect, see \citet{1979rpa..book.....R}; we discuss this issue below.

We convolve $\mathcal{T}_{\rm sd}$ with the rest-frame reflection model {\tt hreflect} \citepalias{2019MNRAS.485.2942N} to find the radius-dependent reflection spectrum. {\tt hreflect} is the hybrid model, using {\tt xillver} \citep{2013ApJ...768..146G} in the soft X-ray range and {\tt ireflect} \citep{1995MNRAS.273..837M} in the hard X-ray range, and applying correction on the temperature parameter in {\tt xillver} to account for inaccuracies of {\tt nthcomp} applied in {\tt xillver}. We convolve $\mathcal{T}_{\rm so}$ with the Comptonization model {\tt compps} \citep{1996ApJ...470..249P} to find the spectrum of the directly observed radiation.  The model is parameterized by the electron temperature, $T_{\rm e}$, measured in the rest-frame of the X-ray source.
The inclination angle of a distant observer is denoted by $i$.

We take into account the spatial extent of the X-ray source. We assume that it has  a cylindrical shape with radius $r_{\rm c}$ and is located symmetrically around the BH rotation axis between a lower height $h_{\rm min}$ and upper height $h_{\rm max}$. Setting $\Delta h \ll h$ and $r_{\rm c} \ll h$ reproduces spectra for a point-like source, where  $\Delta h  = h_{\rm max} - h_{\rm min}$ and $h  = (h_{\rm min} + h_{\rm max})/2$. The \texttt{reflkerrV} spectra presented in this work are computed for $\Delta h  = 0.2$ and $r_{\rm c} = 0.2$ (which in all cases gives a precise imitation of a point-like source (to better than 0.1 per cent) and for brevity we only give the values of $h$; the assumption of $r_{\rm c} = 0.2$ is released only when we find the upper limit on $r_{\rm c}$ in the model of \citetalias{2015ApJ...808....9P} in Section \ref{sect:cygx1}.
All length scales $r,h$ are in units of the gravitational radius, $R_{\rm g} = GM/c^2$. 

 \begin{figure}
\centering
 \includegraphics[width=6cm]{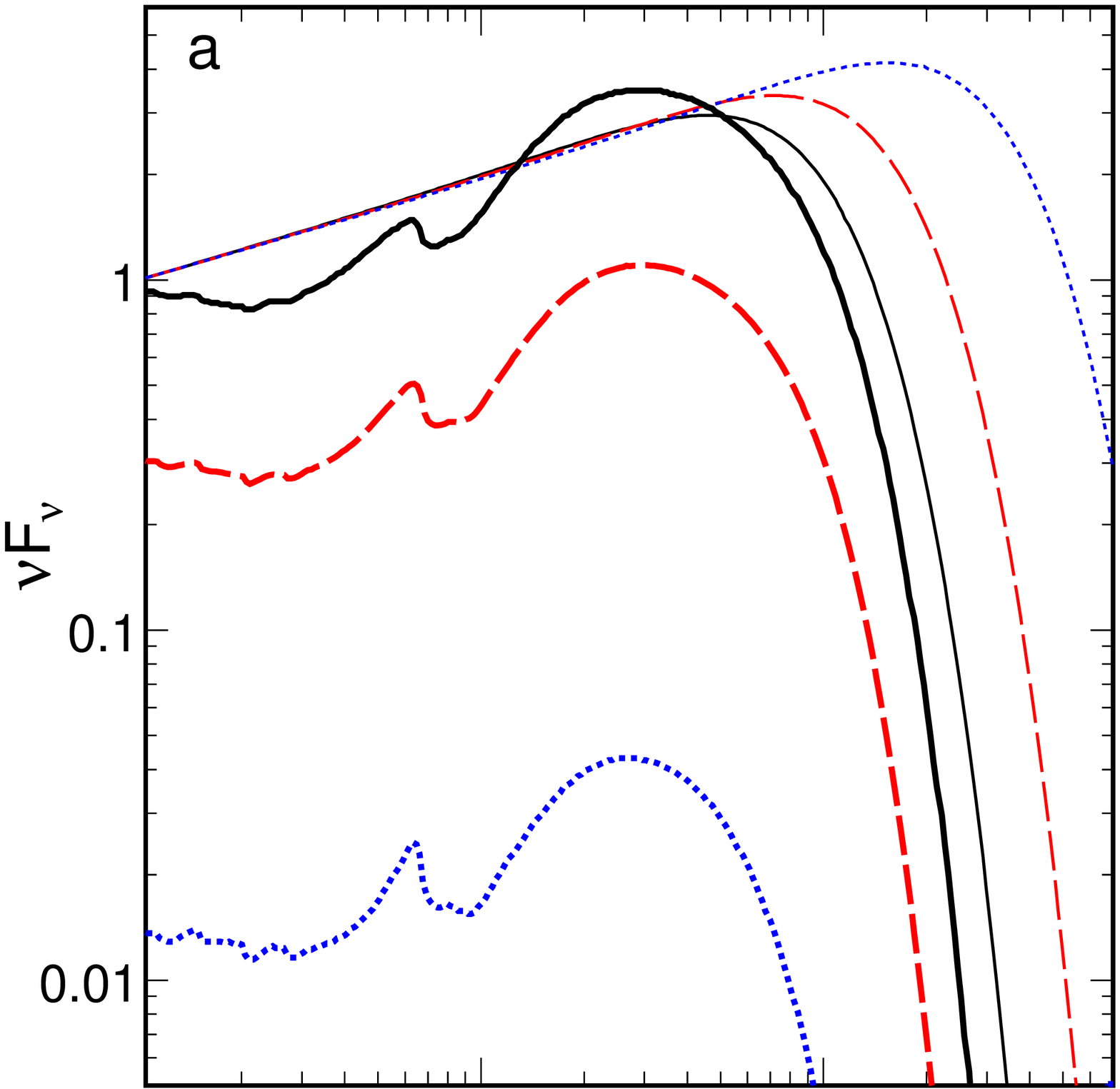}
 \includegraphics[width=6cm]{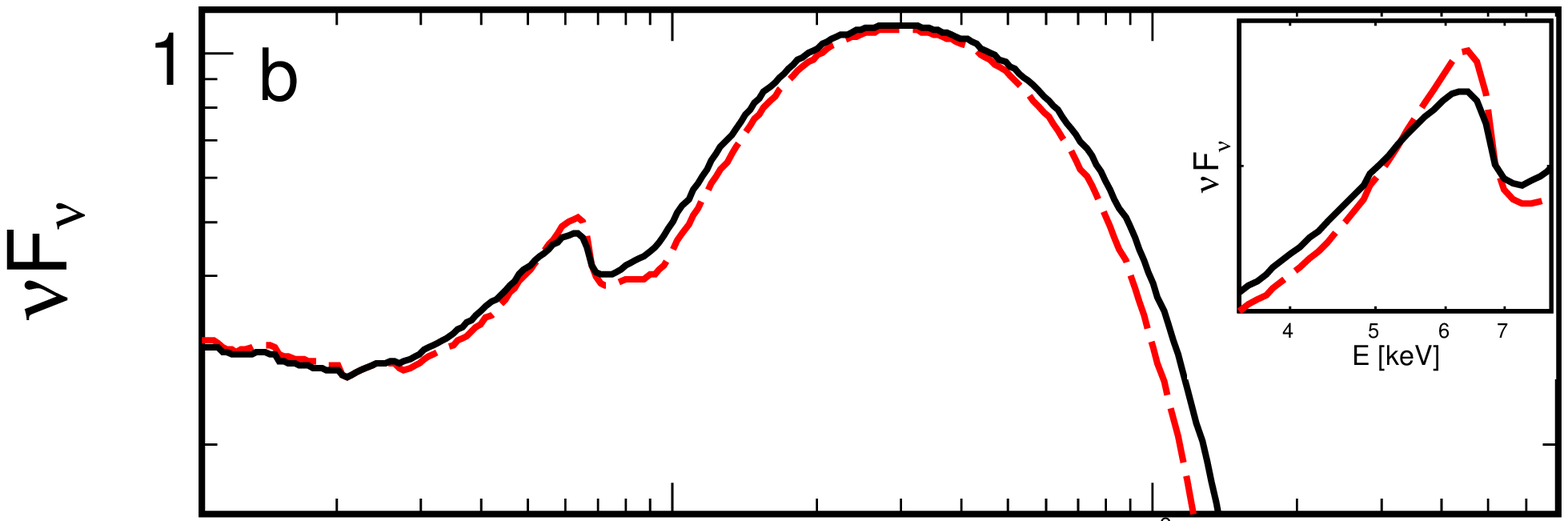}
 \includegraphics[width=6cm]{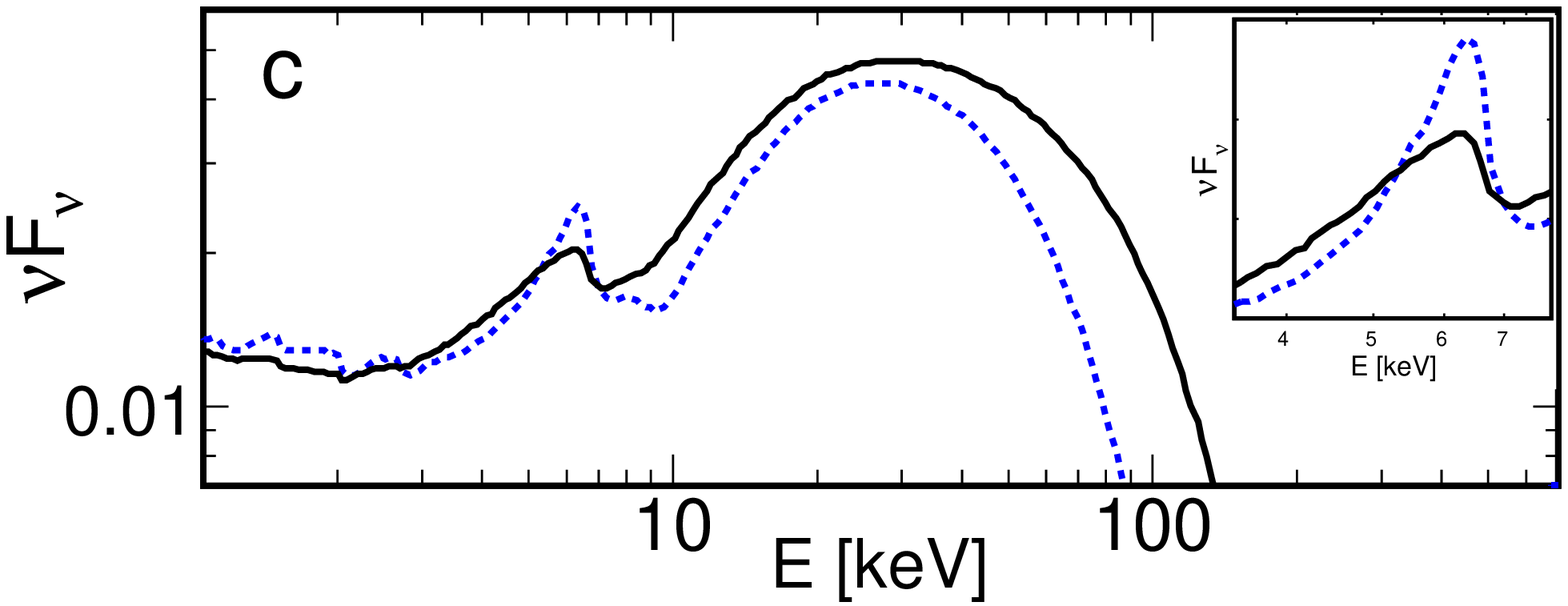}
\caption{(a) Observed primary and reflected spectra for $a=0.998$, $r_{\rm in} = r_{\rm ISCO}$, $h=3$, $i = 20 \degr$, $\Gamma=1.7$, $kT_{\rm e} = 40$ keV and $\beta=0$ (solid black), 0.44 (dashed red) and 0.88 (dotted blue), computed using {\tt reflkerrV}. The ionization parameter $\xi = 10^3$ and the relative iron abundance  $Z_{\rm Fe} = 1$. Panels (b) and (c) show the same reflection spectra for $\beta=0.44$ and $0.88$, respectively, together with the reflection spectrum for $\beta=0$ rescaled by {\tt rel\_refl}~=~0.32 in (b) and by {\tt rel\_refl}~=~0.014 in (c); the inner panels show the same focused on the Fe K$\alpha$ line.  The difference between spectra shown in the bottom panels gives the inaccuracy related with setting  the  reflection scaling parameter to values different from unity.   
}
\label{fig:vel}
\end{figure}

The outward motion of radiating electrons affects the observed spectrum through several effects: (i) weaker relativistic broadening due to reduced irradiation of the innermost disc; (ii) reduction of the incident flux due to the Doppler redshift and collimation away from the disc, (iii) decrease or increase (depending on $i$) of the directly received (i.e.\ not reflected)  flux due to the Doppler shift and collimation; (iv) decrease of the high-energy cut-off of the incident spectrum due to the Doppler redshift; and (v) Doppler shift of the high-energy cut-off of the directly observed spectrum. The effect (iv) affects both the reflected Compton hump, whose shape is sensitive to the position of the high-energy cut-off \citep{1995MNRAS.273..837M} and the soft X-ray part of the reflected spectrum \citep{2015ApJ...808L..37G}. For effects (iv) and (v) we note that it is important to properly shift in energy the precise Comptonization spectrum rather than to apply scaling of temperature in a non-relativistic Comptonization model, e.g.\ in {\tt nthcomp} \citep{1996MNRAS.283..193Z} used in {\tt relxill} and in {\tt reltrans}. The latter approach may lead to nonphysical results if the implied rest-frame temperature is relativistic, see e.g.\ \citet{2021ApJ...909..205S}. 

Fig.\ \ref{fig:vel} illustrates these effects by showing changes of the reflection spectra corresponding to the increase of $v$. For face-on observers, this leads to increasing reduction of the reflected component. However, the decrease in relativistic broadening for increasing $v$ is also clear, as is the change in the shape of the Compton hump.

\begin{figure}
\centering
 \includegraphics[width=4.41cm]{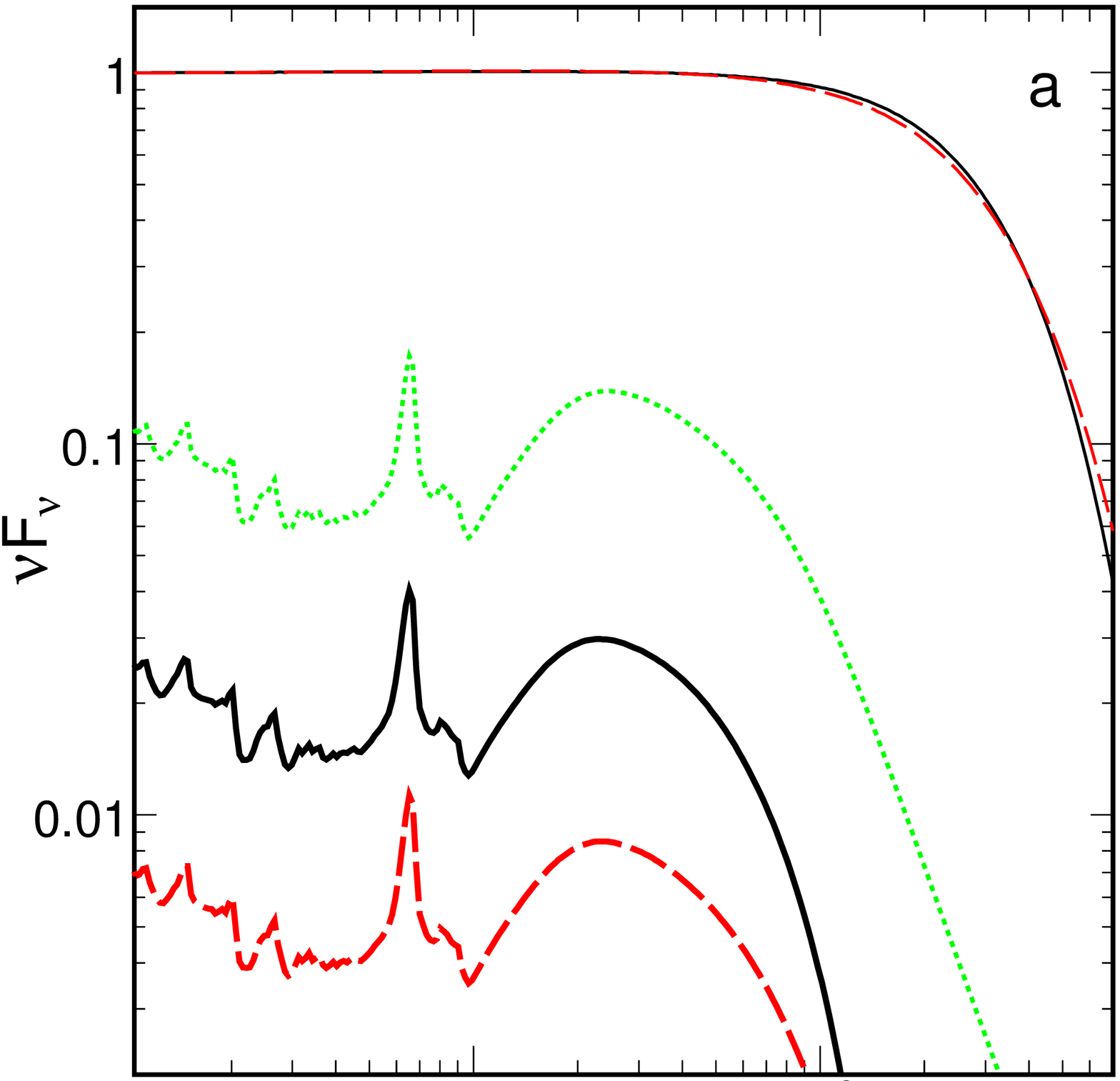}\includegraphics[width=3.9cm]{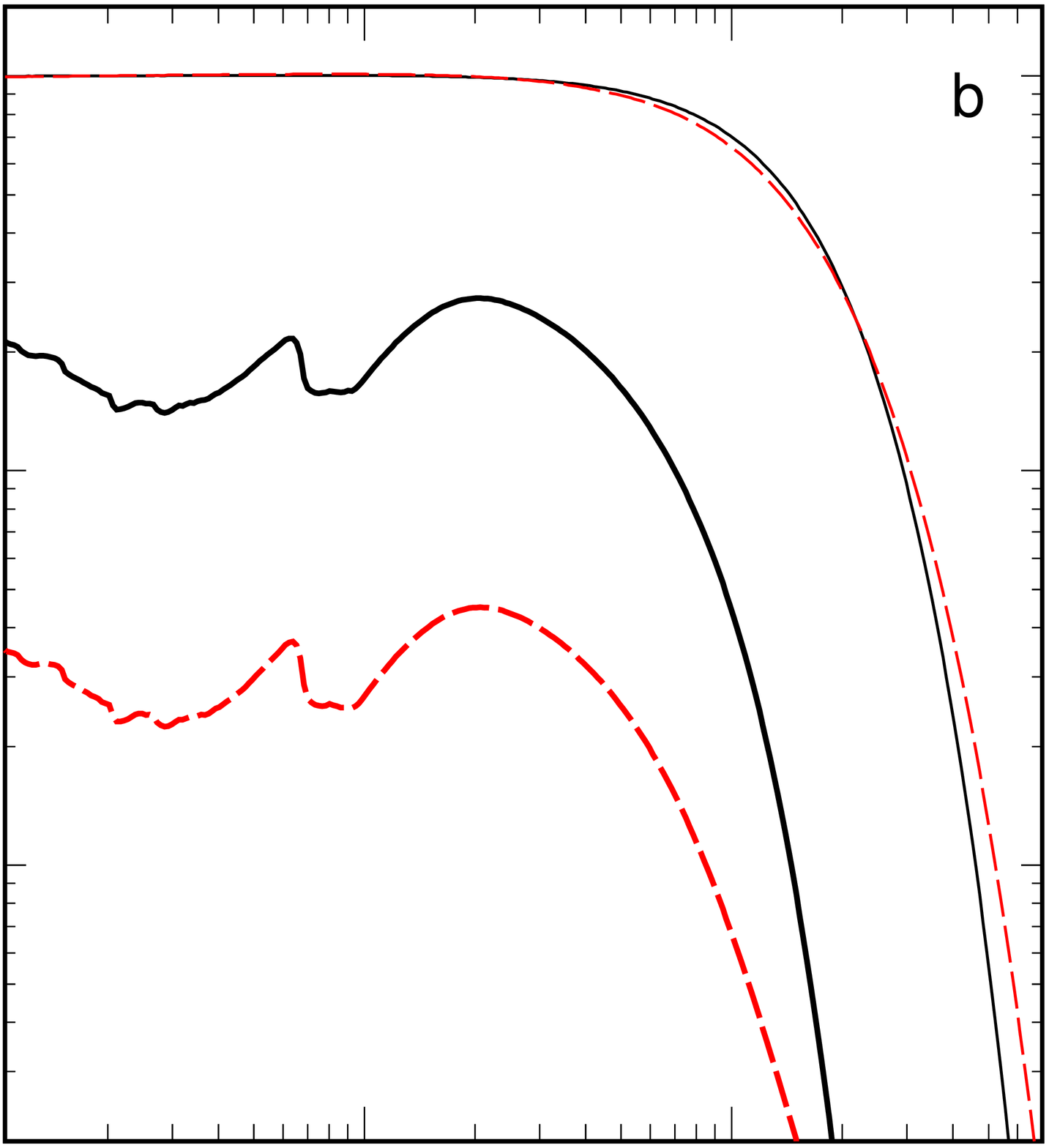}\\
 \includegraphics[width=4.41cm]{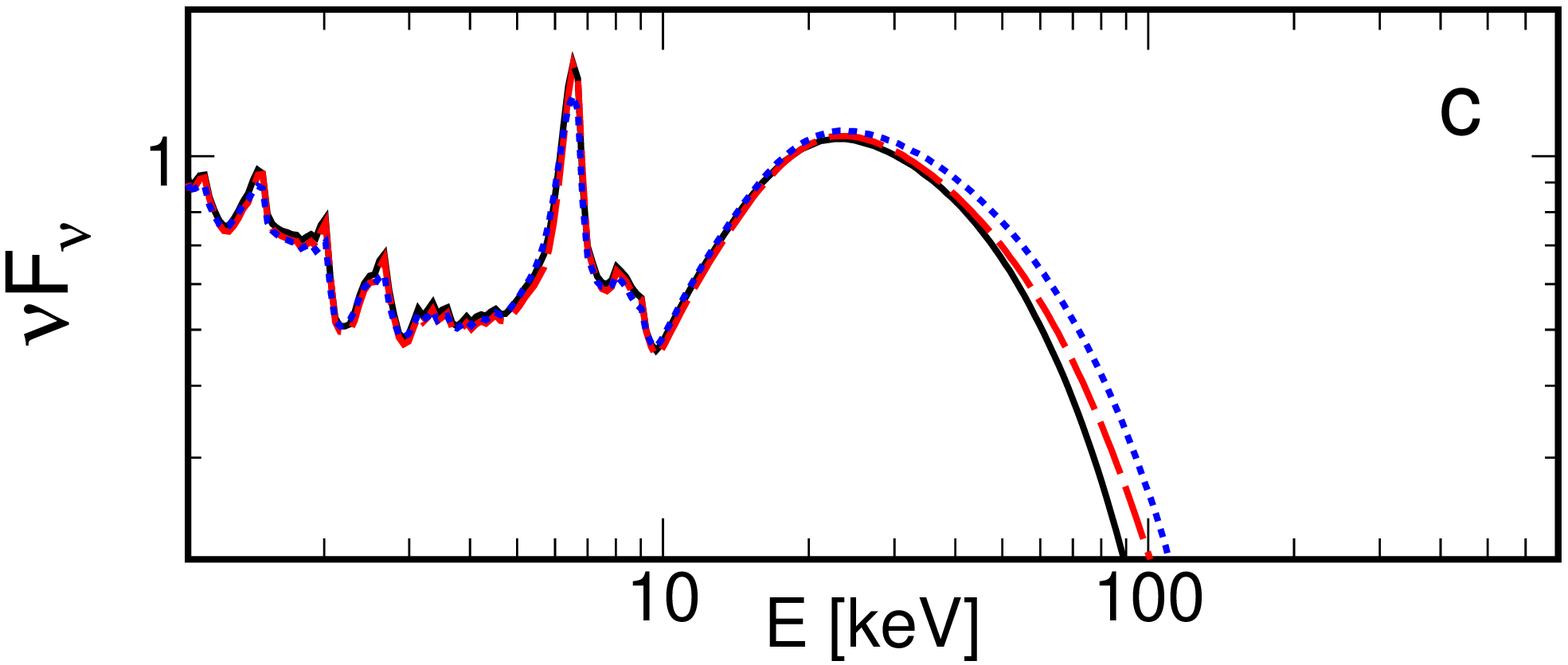}\includegraphics[width=3.9cm]{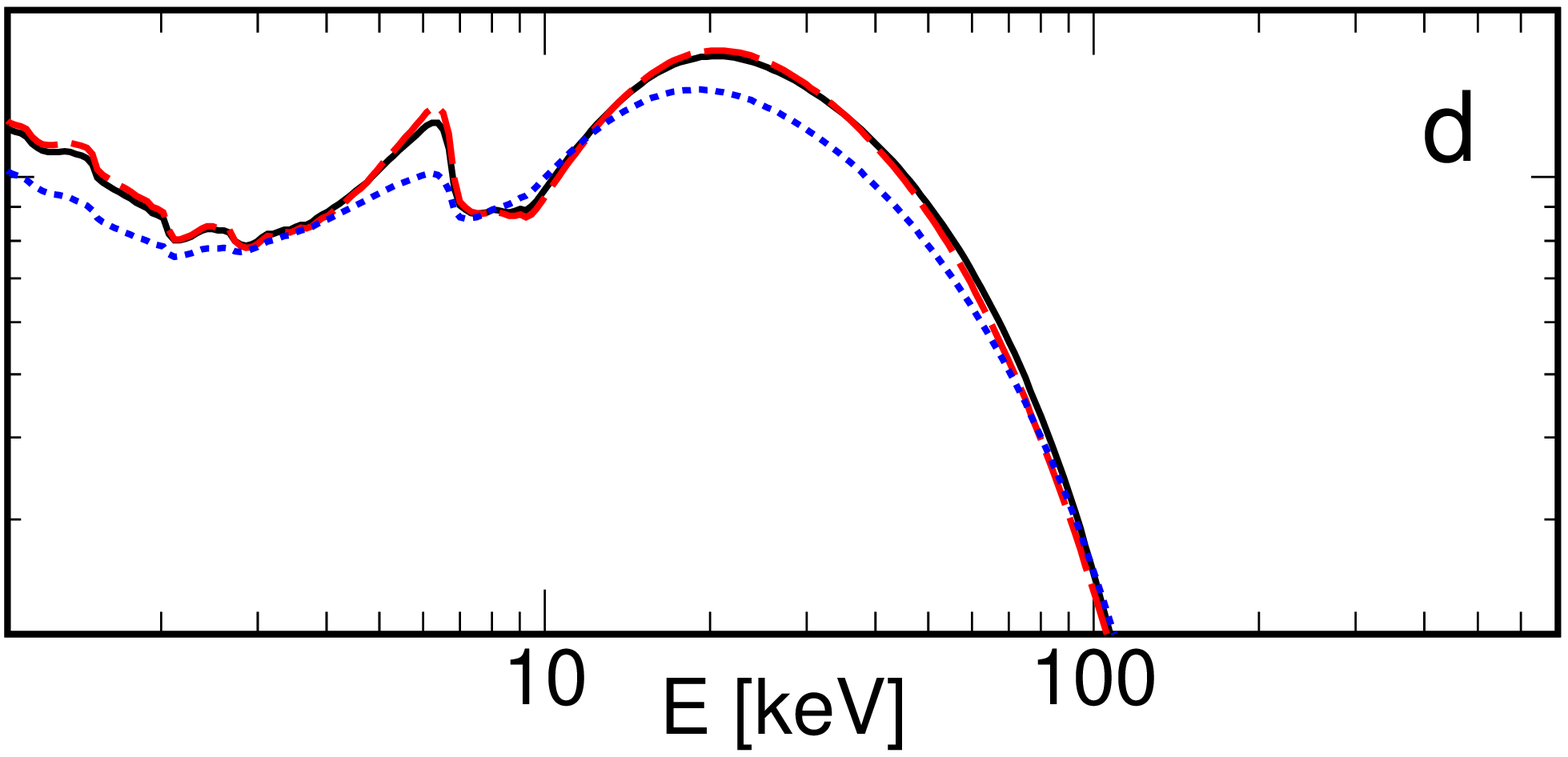}
\caption{Comparison of the observed spectra computed with  {\tt reflkerrV} (solid black), {\tt relxilllpCp}  v.\ 2.1 (dashed red) and v.\ 2.0 (dotted green in panel a) for $\beta=0.66$, $a=0.998$, $r_{\rm in} = r_{\rm ISCO}$,  $\theta = 20 \degr$, $\Gamma=2$, $\xi = 10^3$, $Z_{\rm Fe} = 1$, (a,c) $h=30$ and (b,d) $h=2$. In {\tt reflkerrV}, the (rest-frame) $kT_{\rm e} = 50$ keV. In {\tt relxilllpCp},   $kT_{\rm e} = 100$ keV, for which its primary spectrum matches that of {\tt reflkerrV} (compensating for the shortcomings of {\tt nthcomp}, see \citetalias{2019MNRAS.485.2942N}). Panels (a,b) show spectra with the physical normalization of reflection. 
Panels (c,d) show the same spectra with the  reflected components rescaled by (c) {\tt rel\_refl}~=~36 in {\tt reflkerrV} and {\tt boost}~=~130 in {\tt relxilllpCp} and (d) {\tt rel\_refl}~=~5.6 in {\tt reflkerrV} and {\tt boost}~=~35 in {\tt relxilllpCp}. 
The bottom panels show also the {\tt reflkerrV} spectra for $\beta=0$ and the same remaining parameters as the top panels, in particular $kT_{\rm e} = 50$ keV, and the physical normalization of reflection (blue dotted). }
\label{fig:models}
\end{figure}

All spectra and fitting results presented in this work, except for Fig.\ \ref{fig:vel}bc, Fig.\ \ref{fig:models}cd and model 1 in Section \ref{sect:cygx1}, correspond to the physical normalization of the reflected component, i.e.\ {\tt boost}~=~1 in {\tt relxilllpCp} and {\tt rel\_refl}~=~1 in {\tt reflkerrV}.
Allowing for a free normalization of reflection with respect to the directly observed component, which has become a common practice in applications of the LP model, takes into account only the above effects (ii) and (iii), neglecting the remaining ones. 
Panels (b) and (c) in Fig.\ \ref{fig:vel} illustrate inaccuracy related with approximating the non-isotropy of the X-ray source by simple rescaling of the reflection component.

In Fig.\ \ref{fig:models} we compare spectra computed with {\tt reflkerrV} and {\tt relxilllpCp}. We note that the outflow effect was incorrectly implemented in version 2.0 and earlier versions of  {\tt relxilllpCp}. The corrected version 2.1 gives a systematically lower reflection strength than {\tt reflkerrV}, which appears to be due different physical assumptions in these models, as discussed below.

We first estimate the factor by which the vertical motion decreases the amplitude of reflection. We denote by $\mathcal{F}$ the ratio of the energy flux directly reaching the observer to the luminosity intercepted by the disc. For a power-law spectrum with the photon spectral index $\Gamma$, we have $\mathcal{F}(\beta) \simeq \mathcal{F}(0)\mathcal{A}^{-1} \delta_{\rm obs}^2 (\delta_{\rm obs}/\delta_{\rm eff})^{\Gamma - 1} $, where the last term gives the change of the observed and incident fluxes due to the shift in energy, $\delta_{\rm obs}^2$ is the aberration factor for the directly received photon flux, $\mathcal{A}$ denotes the reduction of the photon flux emitted toward the disc, $\delta_{\rm obs} = 1/[\gamma (1 - \beta \mu)]$,  $\mu = \cos i$, $\gamma = (1 - \beta^2)^{-1/2}$, $\beta = v/c$, $\mathcal{F}(0)$ is for $\beta = 0$  and for the sake of discussion of the involved effects we defined the effective Doppler shift of photons reaching the disc, $\delta_{\rm eff} = \int \delta_{\rm irr}(r) n_{\rm ph}(r) r{\rm d}r/\int n_{\rm ph}(r) r{\rm d}r$, where $\delta_{\rm irr}(r)$ and $n_{\rm ph}(r)$ are the radius-dependent Doppler factor and flux of incident photons. The above expression for $\mathcal{F}(\beta)$ agrees to better than 1 per cent with equation (3) in  \citetalias{1999ApJ...510L.123B}.

Neglecting the light bending, we have $\mathcal{A} = (1 - \beta)$, which follows from $\mu = (\mu' - \beta)/(1 - \beta \mu')$, where $\mu'$ is the cosine of the emission angle in the co-moving frame. At $h=30$ and $\beta=0.66$, we get $\delta_{\rm eff}  \simeq 0.6$ and then, for $\Gamma = 2$ and $i=20 \degr$,  
$\mathcal{F} \simeq 38 \mathcal{F}(0)$. Taking into account the (weak) light bending at $h=30$, we get $\mathcal{F}_{\rm GR} \simeq 36 \mathcal{F}_{\rm GR}(0)$. We indeed find that the reflected component is reduced by this factor in {\tt reflkerrV} (Fig.\ \ref{fig:models}a,c).

At low $h$, the beaming effect is less effective in the reduction of reflection, as the light-bending causes both $\delta_{\rm eff}$ and $\mathcal{A}$ to be larger than at large $h$. E.g.\ at $h=2$ and $v=0.66$, $\delta_{\rm eff} \simeq 1$ and $\mathcal{A} \simeq 0.9$ (the flux of photons directly observed increases at the expense of the flux of photons trapped by the BH, however, the total flux of photons incident on the disc is weakly affected). Then, for $i=20 \degr$ and $\Gamma=2$,  $\mathcal{F}_{\rm GR} \simeq 10 \mathcal{F}_{\rm GR}(0)$. Moreover, while the total irradiating flux changes weakly, its radial distribution changes significantly; the photon flux irradiating the disc at $r < 2$ ($r>6$) decreases (increases) by a factor of $\simeq 2$. As a result, the reflected component computed with \texttt{reflkerrV} for $i=20 \degr$ and $\beta = 0.66$ is reduced by only a factor of $\simeq 0.18$ compared to $\beta=0$ (see Fig.\ \ref{fig:models}b,d), as reflection from $r < 2$ contributes weakly to the observed reflection.

In the above, we neglected the retardation terms, because the emission region is located at a fixed $h$ in our LP model. For an emission region receding from the disc, the retardation effect would have to be taken into account. Then, $\mathcal{F(\beta)}/\mathcal{F}(0)$ would be increased by $\delta_{\rm obs}/\delta_{\rm eff}$ at large $h$ (i.e.\ neglecting GR), which gives the factor of $\simeq 3.3$ for the case shown in Fig.\ \ref{fig:models}a, explaining the difference between \texttt{reflkerrV} and {\tt relxilllpCp} v.\ 2.1. At this large $h$ the retardation effect does not affect the spectrum of the reflected component, which arises from a range of radii corresponding to a narrow range of $\delta_{\rm irr} \approx \delta_{\rm eff}$. We indeed note a good agreement of the spectral shape of reflection computed with \texttt{reflkerrV} and {\tt relxilllpCp}, see Fig.\ \ref{fig:models}c. At low $h$, the retardation changes the relative contribution of photons reflected at different radii, which slightly affects the spectral shape, e.g.\ leading to $\la 10$ per cent differences  between the \texttt{reflkerrV} and {\tt relxilllpCp} spectra in Fig.\ \ref{fig:models}d.

\begin{table}
 \caption{Results of spectral fitting for our models with stratified Comptonization. All the models are defined as {\tt tbabs*wind*(LP + compps + diskbb)}, where {\tt LP} is described by {\tt reflkerrV} in models 4 and 5, and {\tt relfkerrG\_lp} in model 6.
 }
\begin{center}
\begin{tabular}{llll}
 \hline
 & 4 & 5 & 6    \\
 \hline
 & \multicolumn{3}{c}{LP component} \\
 $h$                    & 
$16.5^{+0.8}_{-1.6}$      & $6.2^{+0.4}_{-2.2}$       & $11.8^{+0.9}_{-3.1}$\\[0.1cm]
 $a$                & 
$0.998^{+0}_{-0.998}$        & $0.998$(f)       & $0.998$(f)\\[0.1cm]
 $\beta$                &  
$0.36^{+0.03}_{-0.01}$        & $0.18^{+0.07}_{-0.07}$       & $-$\\[0.1cm]
 $\delta_{\rm B}$ & $-$            & 
$-$     & $>0.5$\\[0.1cm]
 $r_{\rm in} ~~ [R_{\rm g}]$ & 
$r_{\rm ISCO}$(f)      & $16.0^{+1.8}_{-2.0}$     & 
$15.7^{+3.2}_{-2.4}$  \\[0.1cm]
 $i \; [^{\circ}]$      & 
$41.7^{+0.8}_{-1.0}$        & $39.6^{+1.7}_{-1.6}$       & $41.1^{+2.9}_{-1.9}$\\[0.1cm]
 $\Gamma$               & 
$1.71^{+0.01}_{-0.003}$      & $1.72^{+0.01}_{-0.01}$     & $1.72^{+0.01}_{-0.01}$\\[0.1cm]
 $kT_{e} ~~ [{\rm keV}]$               & 
$46^{+2}_{-1}$        & $64^{+5}_{-6}$       & $69^{+2}_{-2}$\\[0.1cm]
 $Z_{\rm Fe}$           & 
$2.0^{+0.1}_{-0.1}$        & $2.1^{+0.2}_{-0.3}$       & $2.1^{+0.1}_{-0.2}$\\[0.1cm]
 $\log_{10}(\xi) $      & 
$2.86^{+0.02}_{-0.02}$        & $2.88^{+0.04}_{-0.03}$       & $2.86^{+0.03}_{-0.03}$\\[0.1cm]
 $N$                    & 
$1.33^{+0.08}_{-0.04}$        & $1.35^{+0.04}_{-0.05}$       & $1.32^{+0.03}_{-0.04}$\\[0.1cm]
 \hline
 & \multicolumn{3}{c}{{\tt tbabs}} \\
 $N_{H} ~~ [\times 10^{22} {\rm cm}^{-2}]$          &  
$0.5^{+0.09}_{-0}$   & $0.5^{+0.2}_{-0}$        & $0.5^{+0.2}_{-0}$\\[0.1cm]
 \hline
 & \multicolumn{3}{c}{ionized absorber: {\tt xstar} table model} \\
 $N_{H} ~~ [\times 10^{22} {\rm cm}^{-2}]$        & 
$2.5^{+0.1}_{-0.1}$   & $2.5^{+0.2}_{-0.2}$  & $2.4^{+0.1}_{-0.2}$\\[0.1cm]
$\log_{10}(\xi) $      & 
$5.00^{+0}_{-0.07}$        & $5.00^{+0}_{-0.09}$       & $5.00^{+0}_{-0.08}$\\[0.1cm]
 \hline
 & \multicolumn{3}{c}{{\tt diskbb}} \\
 $T_{\rm in} ~~ [\times 10^{-2} {\rm keV}]$        & 
$9.1^{+0.7}_{-1.9}$      & $9.5^{+3.3}_{-2.5}$    & $9.4^{+2.9}_{-2.3}$\\[0.1cm]
 $N ~~ [\times 10^{8}]$                & 
$4.5^{+13}_{-2.6}$    & $2.8^{+34}_{-2.7}$  & $2.9^{+33}_{-1.7}$\\[0.1cm]
 \hline
 & \multicolumn{3}{c}{soft component: {\tt compps}} \\[0.05cm]
 $\Gamma$        & 
$2.97^{+0.25}_{-0.09}$      & $3.05^{+0.37}_{-0.16}$       & $2.91^{+0.36}_{-0.05}$\\[0.1cm]
 $kT_{e}$        & 
$7^{+1}_{-1}$      & $7^{+28}_{-1}$   & $6^{+1}_{-1}$\\[0.1cm]
 $N$        & 
$1.45^{+0.31}_{-0.08}$      & $1.57^{+0.93}_{-0.33}$       & $1.39^{+0.05}_{-0.04}$\\[0.1cm]
 \hline
 $\chi^{2}/{\rm DoF}$   & 
$641/490$              & $627/490$                & $618/490$\\
\end{tabular}\\
\end{center}
{\it Notes:} We use the same model of ionized absorber which was used by \citetalias{2015ApJ...808....9P} and \citetalias{2017MNRAS.472.4220B}. The ionization parameter, $\xi$, is given in the unit of ${\rm erg\; cm\; s}^{-1}$. The normalization, $N$, of Comptonization components gives the 1-keV flux in keV cm$^{-2}$ s$^{-1}$. For the soft component we used the version of {\tt compps} parametrized by $\Gamma$, see \citetalias{2019MNRAS.485.2942N}. 
(f) denotes a fixed parameter.
 \label{tab:fit}
\end{table}

 \begin{figure}
\centering
 \includegraphics[width=7cm]{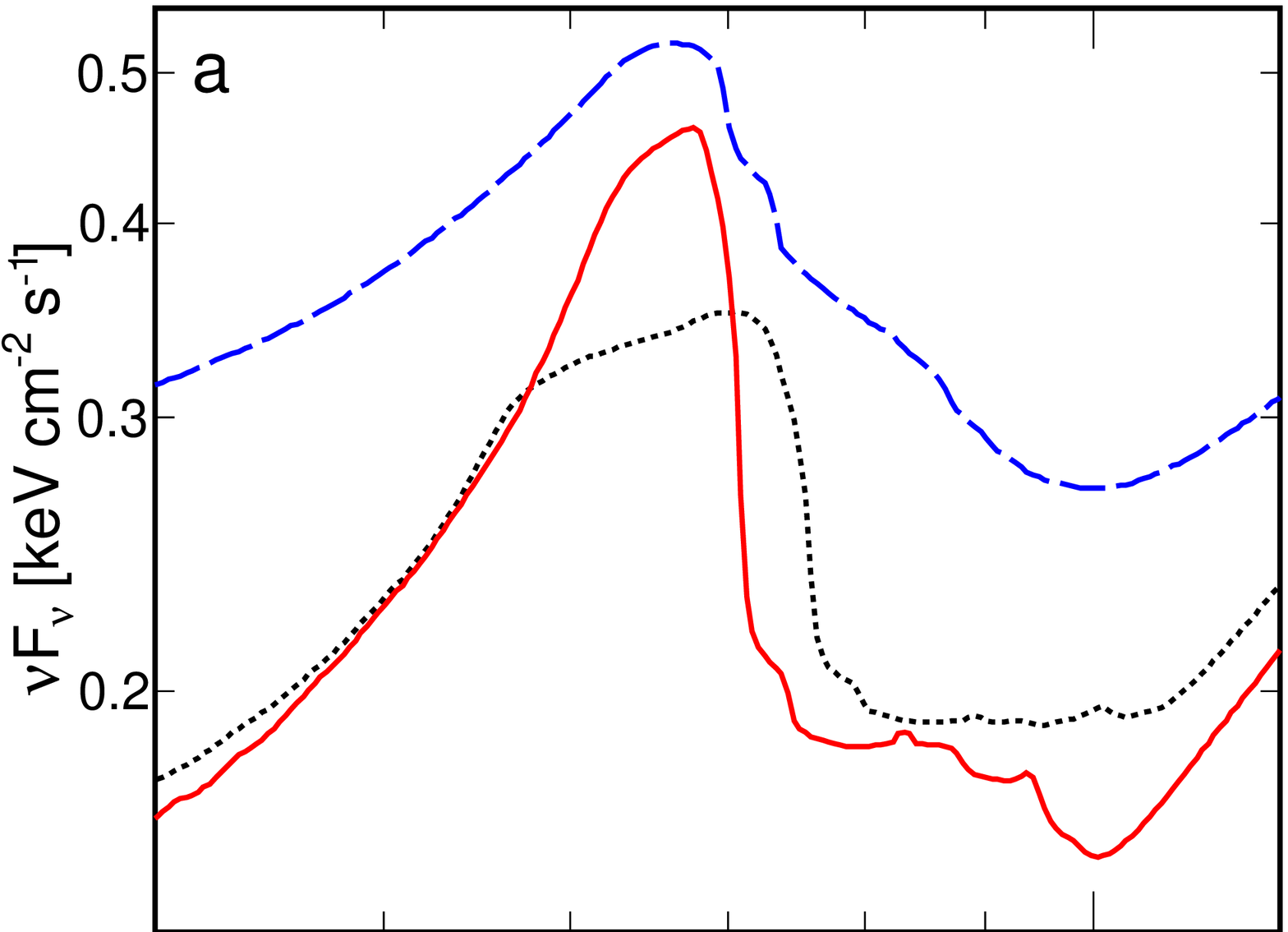}
 \includegraphics[width=7cm]{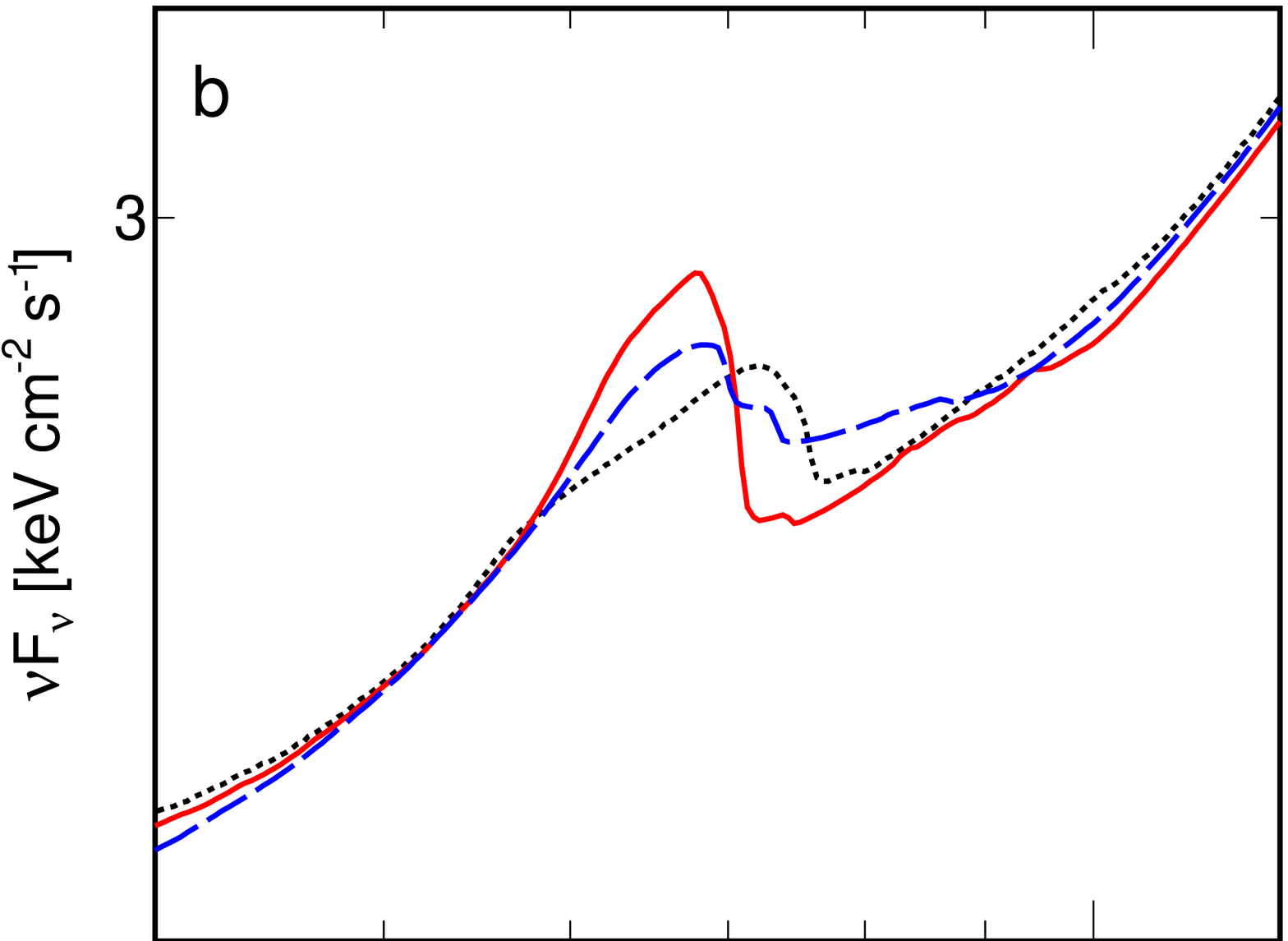}
 \includegraphics[width=7cm]{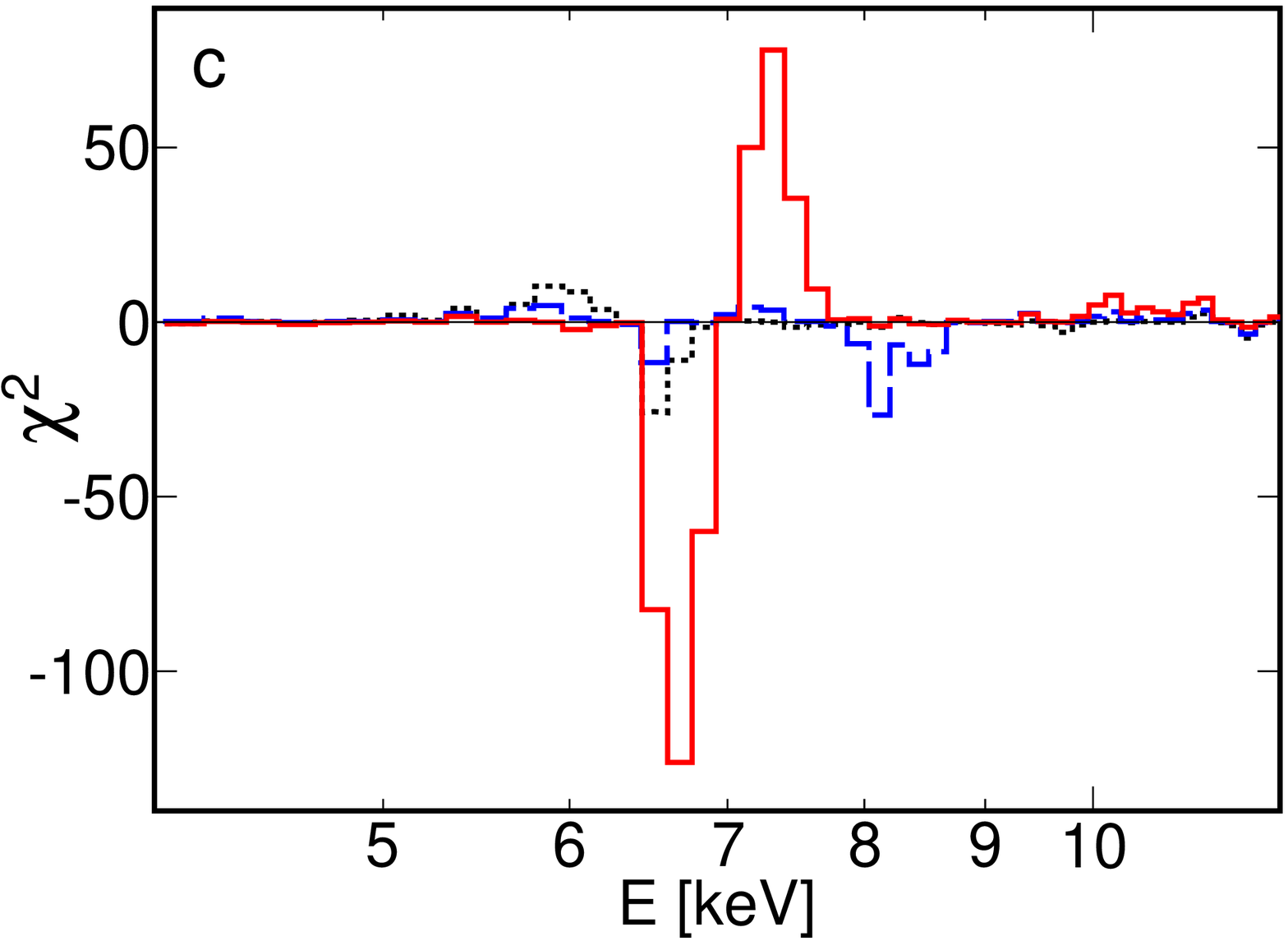}
\caption{Illustration of spectral effects corresponding to using either an outflow or a free scaling of reflection in the model of \citetalias{2015ApJ...808....9P}, see text. In all panels, the black dotted lines are for model 1 (with $\beta=0$ and free $\mathcal{R}$), the red solid lines are for model 2 and the blue dashed lines are for model 3. Models 1 and 2 use the same parameters of {\tt eqpair}, $Z_{\rm Fe}$ and $\xi$, so the difference between the solid and dotted lines directly reflects the spectral differences shown in Figure \ref{fig:vel}bc. a) Reflection components. b) Total spectra. c) Fit residuals given as $\chi^2$ contributions for {\it NuSTAR}; the FPMA and FPMB data are co-added for clarity of this figure.     
}
\label{fig:p15}
\end{figure}

\begin{figure*}
\includegraphics[height=4.99cm]{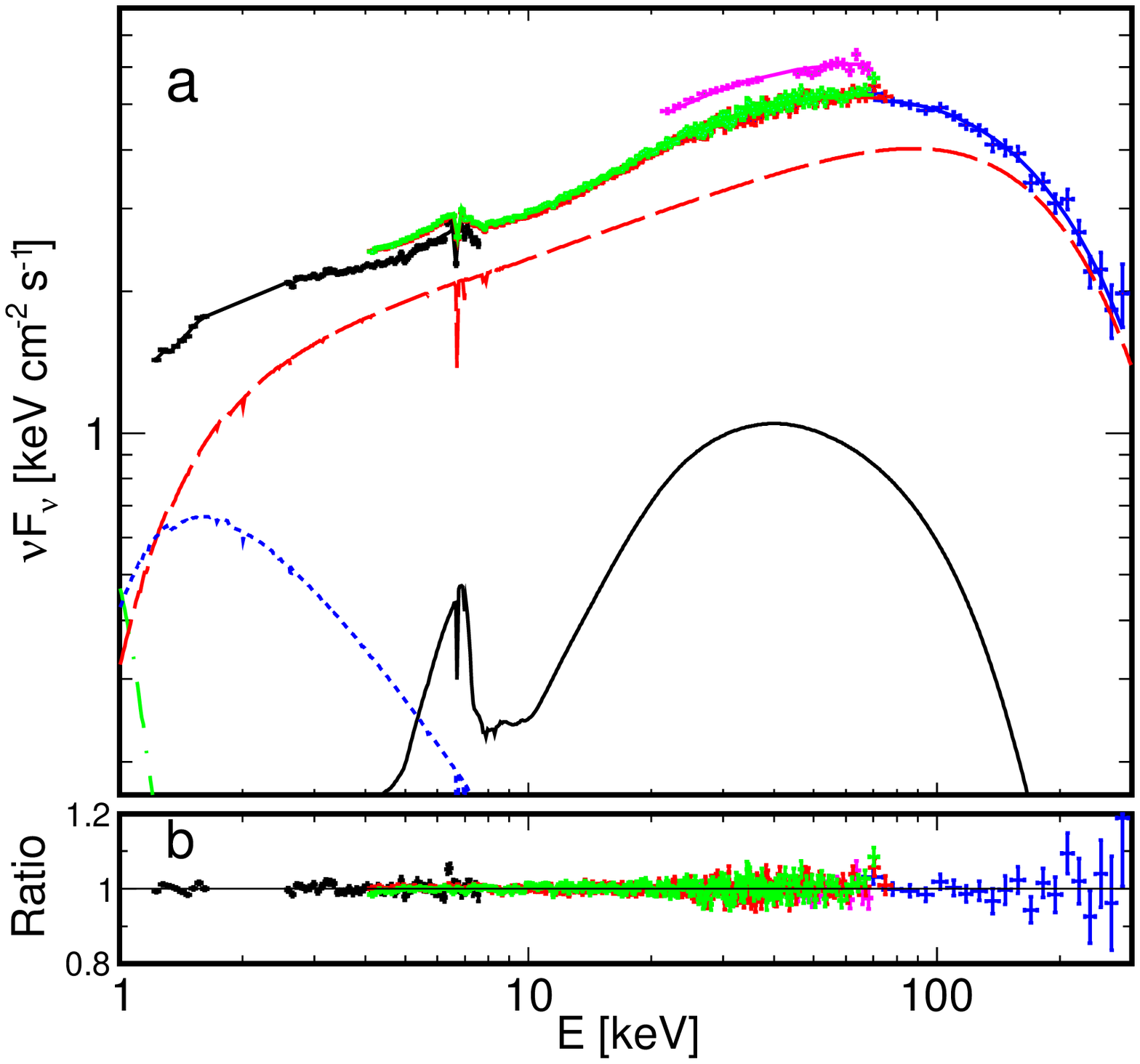}
\hspace{0cm}
\includegraphics[height=4.99cm]{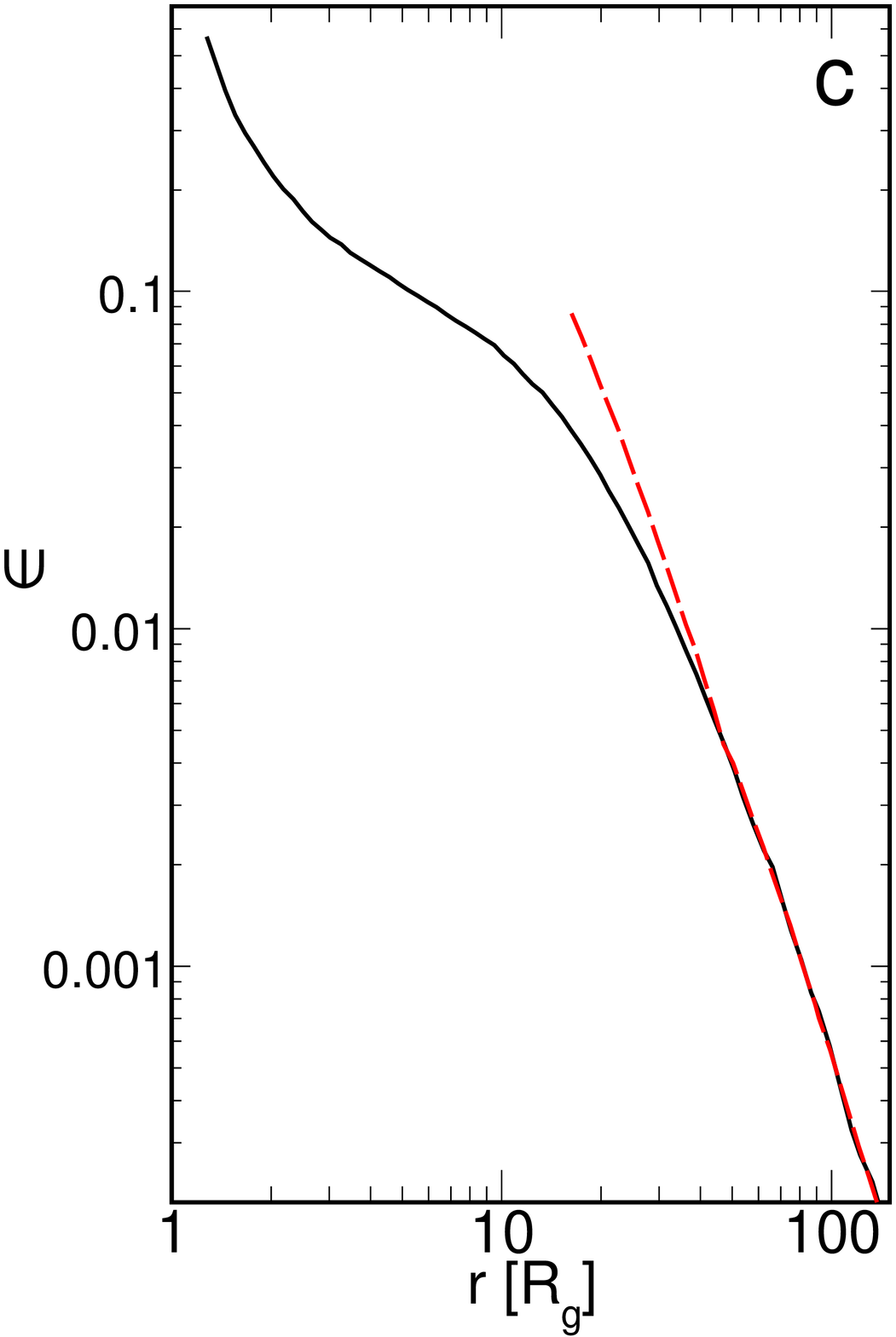}
\hspace{0.5cm}
\includegraphics[height=4.99cm]{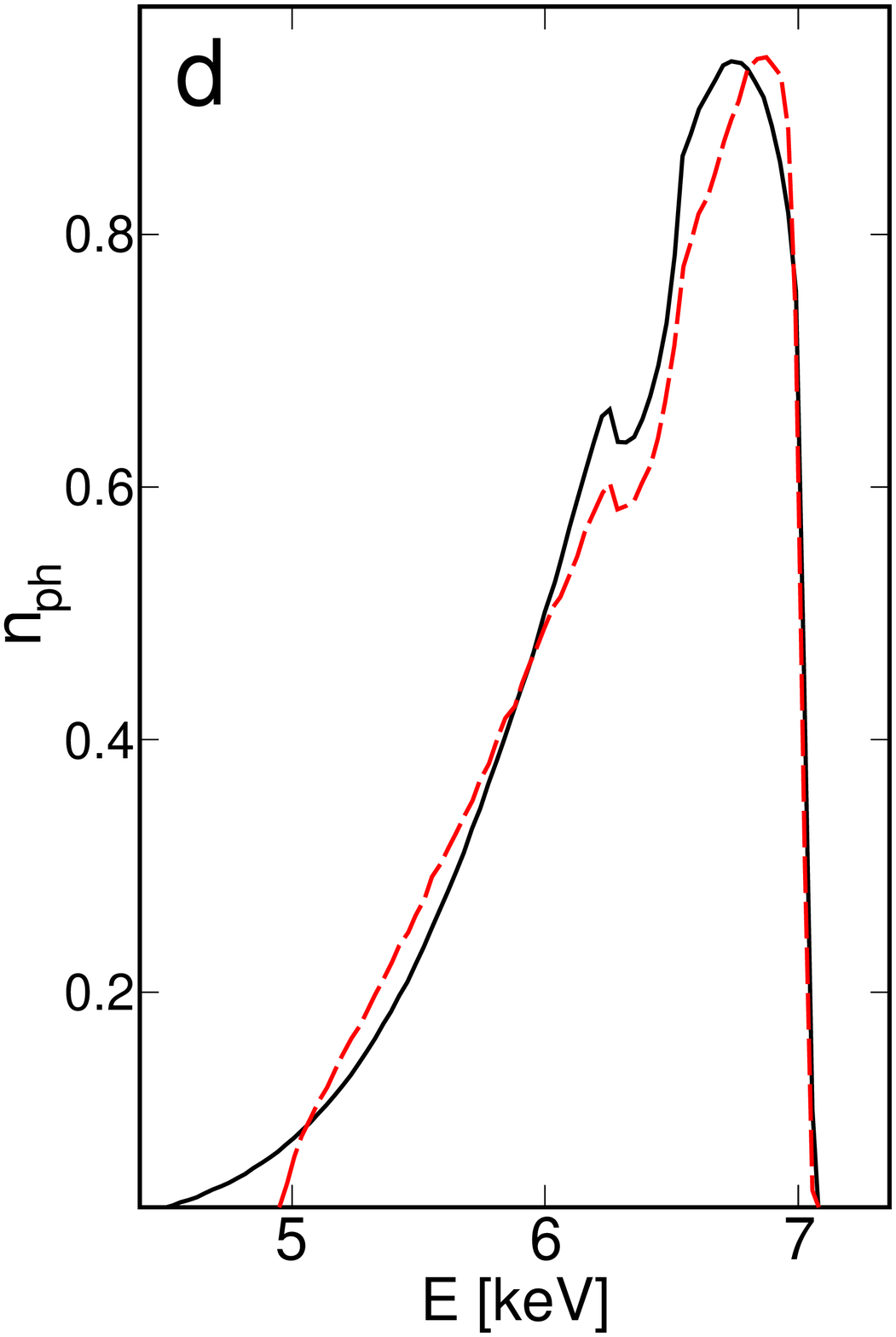}
\hspace{0.5cm}
\includegraphics[height=4.99cm]{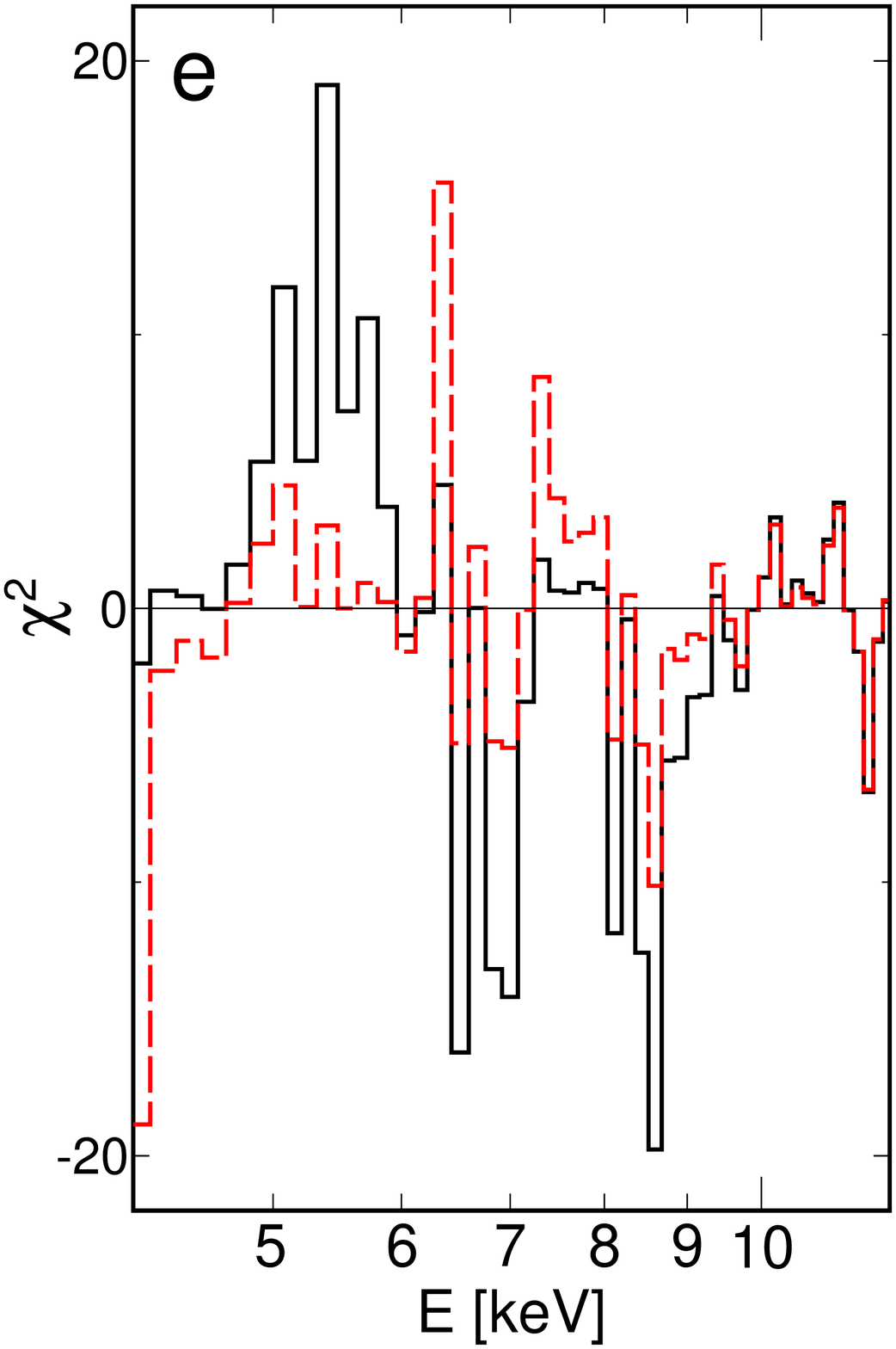}
\caption{(a) Unfolded data and model spectrum for our model 6, see Table \ref{tab:fit}. The data points for FPMA, FPMB, XIS1, PIN and GSO are shown in red, green, black, magenta and blue, respectively. The red dashed line shows the Comptonization component of  {\tt reflkerrV} and the black solid line shows its reflection. The blue dotted line shows the disc blackbody and the green dot-dashed line shows the soft Comptonization. (b) Fit residuals given as the data-to-model ratio. For models 4 and 5, the model components and the data-to-model ratios are very similar to these shown in (a,b). Panels (c-e) illustrate the difference between model 4 (solid black) and model 6 (red dashed).  (c) Radial emissivity profiles of reflection. (d) Profiles of an intrinsically narrow line with the rest-frame energy 6.4 keV, observed at $i = 41 \degr$. (e) Fit residuals given as $\chi^2$ contributions for {\it NuSTAR}; the FPMA and FPMB data are co-added for clarity of this figure. }
\label{fig:cygx1}
\end{figure*}

Models assuming a motion of the source itself (rather than a motion within it) should be applied using a vertically extended emission region with a large $h_{\rm max}$, rather than a point-like source, because typical observation times exceed the time-scale for mildly-relativistic propagation beyond $100 R_{\rm g}$ even for supermassive BHs.

Finally we note also that the reduction of reflection strength may be due to the intrinsic non-isotropy of the X-ray source instead of the kinematic collimation. The above remarks remain valid for such a case, i.e.\ the change of the radial profile corresponding to this non-isotropy should also be taken into account. \texttt{reflkerrV} can be used to self-consistently approximate these alternative non-isotropy effects, as the retardation obviously should not be included for such an approximation. 
A particularly relevant effect concerns the non-isotropy of the Comptonization process. This effect is implemented in the {\tt compps} model and then can be accurately taken into account in spectral modeling. We note, however, that this non-isotropy is relatively weak and leads to changes of the reflection strength by at most a factor of $\sim 2$ (see e.g.\ figure A1 in \citetalias{2019MNRAS.485.2942N}).

\section{Hard-state of Cyg X-1}
\label{sect:cygx1}

We study here the  {\it NuSTAR} and {\it Suzaku} observations of Cyg X-1 in its hard state. We consider the {\it NuSTAR} observation on 2014 May 20-21, OBSID 30001011007, and the  {\it Suzaku} observation on 2014 May 19-22, OBSID 409049010, of which we use only the data simultaneous with the {\it NuSTAR}. We consider data from the FPMA and FPMB detectors onboard {\it NuSTAR} and  from three {\it Suzaku} instruments, XIS1, PIN and GSO. 
 In the data reduction, we followed the standard procedures, similarly to
\citetalias{2015ApJ...808....9P} and \citetalias{2017MNRAS.472.4220B}, and we found the data of all the detectors to be
very similar to those obtained by these works.
For {\it Suzaku} we used the standard reprocessing tool, \texttt{aepipeline} ver.~1.1.0, with its default settings and \texttt{CALDB} ver.\ 2016-06-07, and we generated response files with \texttt{xisrmfgen} and \texttt{xissimarfgen} ftools. 
The reduction of {\it NuSTAR} data   was performed with \texttt{nupipeline} ftool and \texttt{nustardas} v1.4.1 from 2014-05-28 with standard settings and CALDB version files from 2015-03-16. 
All spectra were rebinned using the optimal binning of \citet{2016A&A...587A.151K}

We have refitted the LP models of \citetalias{2015ApJ...808....9P} and \citetalias{2017MNRAS.472.4220B} to the present data\footnote{to recover their results we used the same reflection models as originally applied in these works, i.e.\ {\tt relxilllp} ver.\ 0.2 for \citetalias{2015ApJ...808....9P} and {\tt relxilllpCp} ver.\ 0.5 for \citetalias{2017MNRAS.472.4220B}} and obtained parameters very similar to those given in Table 4 of  \citetalias{2015ApJ...808....9P} and Table 2 of \citetalias{2017MNRAS.472.4220B} for their model 4, with $\chi^2/{\rm DoF} = 665/488$ and 622/490, respectively.
These models assume a free reflection normalization, which is found to be lower by a factor $\ga 5$ than predicted for the fitted LP parameters. We use here {\tt reflkerrV} to demonstrate how these results are affected when the reduction of reflection is accounted for by beaming effects in a physically consistent model. We construct spectral models similar to those of \citetalias{2015ApJ...808....9P} and \citetalias{2017MNRAS.472.4220B}, namely, we include the disc blackbody component modelled with \texttt{diskbb} \citep{1984PASJ...36..741M}, the interstellar absorption modelled by \texttt{tbabs} with $N_{\rm H} = (5-7) \times 10^{21}$ cm$^{-2}$, and the ionized absorber described by the model based on \texttt{xstar} \citep{2001ApJS..133..221K}.

We first consider the solution of \citetalias{2015ApJ...808....9P} in which the primary continuum component is fully 
described by the  hybrid Comptonization model {\tt eqpair} \citep{1999ASPC..161..375C}. Similarly to \citetalias{2015ApJ...808....9P}, we add to it the relativistic reflection, for which we use {\tt reflkerrV}; we include only the reflection component of this model.
To describe the normalization of the {\tt reflkerrV} reflection relative to the {\tt eqpair} continuum,
we define $\mathcal{R}$ as the ratio of the 10-keV fluxes of the primary component of {\tt reflkerrV} to that of {\tt eqpair}.
Setting $\beta=0$ and allowing for a free $\mathcal{R}$, we find the best-fit similar to that of \citetalias{2015ApJ...808....9P}, with $a = 0.98^{+0.01}_{-0.01}$, $h = 1.3^{+0.1}_{-0.1}$, $r_{\rm in} = 2.3^{+0.1}_{-0.1}$,  $i = 49.7^{+0.5}_{-1.4} \degr$, $Z_{\rm Fe} = 4.2^{+0.1}_{-0.1}$, $\log_{10}(\xi) = 2.98^{+0.02}_{-0.02}$ and  $\mathcal{R} = 0.13^{+0.06}_{-0.01}$ at $\chi^2/{\rm DoF} = 655/487$ (model 1). Then, we set $\mathcal{R} = 1$  (i.e.\ the physical value) and use a free $\beta$. The corresponding spectral changes are shown in Figure \ref{fig:p15}. To illustrate the effect of self-consistent inclusion of the effects of non-isotropy instead of the free scaling of reflection, we first fix  $Z_{\rm Fe}$, $\xi$ and the parameters of the primary component ({\tt eqpair}) at the values found above for the model 1 and we allow only the relativistic-blurring parameters of {\tt reflkerrV} to fit freely. This gives a very poor fit for $a = 0.25^{+0.01}_{-0.01}$,  $h = 1.7^{+0.1}_{-0.1}$,  $r_{\rm in} = 5.1^{+0.1}_{-0.1}$, $\beta \simeq 0.86^{+0.02}_{-0.04}$ and  
$i = 34.5^{+0.1}_{-0.2} \degr$ (model 2) with $\Delta \chi^2 = +459$ with respect to our fit with model 1 and strong systematic residuals in the 6--8 keV range (the solid line in Figure \ref{fig:p15}c). Then, we allow all parameters of the model to fit freely. The change in relativistic blurring for $\beta > 0$ is then partially compensated by the change of $Z_{\rm Fe}$ and $\xi$ and we get $\chi^2/{\rm DoF} = 703/487$ for $a = 0.1^{+0.1}_{-0.1}$, $h = 2.2^{+0.2}_{-0.1}$, $r_{\rm in} = 6^{+1}_{-1}$, $Z_{\rm Fe} \simeq 3.4^{+0.1}_{-0.2}$, $\log_{10}(\xi) \simeq 3.31^{+0.01}_{-0.01}$, $\beta \simeq 0.82^{+0.06}_{-0.12}$ and  $i = 31^{+1}_{-1} \degr$ (model 3). For the sake of the discussion below of the role of Comptonization of reflection in the X-ray corona in the original model of \citetalias{2015ApJ...808....9P} (i.e.\ with $\beta=0$ and free $\mathcal{R}$), we also allowed $r_{\rm c}$ in model 1 to fit freely and we found the upper limit of $r_{\rm c} < 0.9$  \citep[see the discussion of the dependence on $r_{\rm c}$ in][]{2020A&A...641A..89S}.

We then consider the model of \citetalias{2017MNRAS.472.4220B} with stratified Comptonization, i.e.\ we include 
 an additional, soft Comptonization component.
Model definitions and fitting results are shown in Table \ref{tab:fit}. We first assume  $r_{\rm in} = r_{\rm ISCO}$ (model 4) and we find a good fit for $\beta \simeq 0.36$. We then allow $r_{\rm in}$ in {\tt reflkerrV} to vary (model 5). The spin value cannot be constrained, therefore, we fix $a=0.998$. In agreement with \citetalias{2017MNRAS.472.4220B} we find that $r_{\rm in} \simeq 16$ is preferred. Allowing for $r_{\rm in} > r_{\rm ISCO}$ improves the fit and this version represents our best spectral solution with {\tt reflkerrV}.  Reflection from such a truncated disc is still relatively strong and a subrelativistic outflow of the X-ray source at $\beta \simeq 0.2$ is needed to reduce it to the observed level. 

Alternatively, the reduction of reflection may be due to the presence of the X-ray source located on the opposite to observer side of the BH (referred to as the bottom source), which is visible when the optically thick disc is truncated. This source negligibly contributes to the disc irradiation while significantly increasing the directly observed radiation, and thus reducing the reflection strength, see  \citet{2018MNRAS.477.4269N} and \citetalias{2019MNRAS.485.2942N}. To study such a solution, we use the {\tt relfkerrG\_lp} model of \citetalias{2019MNRAS.485.2942N}, with two symmetrically located, static X-ray sources (model 6). In this model, the  attenuation of the bottom source is given by the parameter $\delta_{\rm B}$ ($=1$ for a full contribution of this source). The model 6 gives our overall best spectral solution and we show it in Fig.\ \ref{fig:cygx1}a,b. We note that the parameters fitted in it closely approximate the geometry of a truncated outer disc irradiated by the inner flow with the scale height $h/r \sim 1$, with the two sources representing the parts of the flow on the opposite sides of the equatorial plane. 
 In particular, in such a geometry the fitted $\Gamma$ and $T_{\rm e}$ would correspond to  the optical half-thickness of the flow of $\tau \sim 0.6$, implying attenuation of radiation from the bottom part of the flow by $\sim 50$\%, consistent with the fitted $\delta_{\rm B}$.

Our fitted models indicate that reduced reflection from the central $16R_{\rm g}$ is preferred to fit the data, which may be explained by either an outflow or a disc truncation, see Fig.\ \ref{fig:cygx1}c. The difference in relativistic broadening corresponding to these two cases is small, see Fig.\ \ref{fig:cygx1}d, yet it gives rise to the systematic difference in the Fe K$\alpha$ range in the residuals for the fitted models, see Fig.\ \ref{fig:cygx1}e.

\section{Summary}
\label{sect:sum}

We considered the effect of the vertical outflow at the X-ray source in the popular relativistic-reflection model assuming the LP geometry. We pointed out effects involved in spectral formation and we implemented them in the new model, {\tt reflkerrV}.
We note differences with   {\tt relxilllpCp}, mostly in the reflection strength, related with the neglect of retardation effect in our model corresponding to our assumption of a steady location of the LP source.

{\tt reflkerrV} gives a good description of the X-ray spectrum of Cyg X-1 in its hard state for $\beta=0.36$, which value is in approximate agreement with previous estimations of the outflow velocity from the amount of reflection in this object \citep[e.g.\ \citetalias{1999ApJ...510L.123B};][]{2001MNRAS.326..417M,2001ApJ...547.1024D}. However, a slightly better solution is found in the model involving a truncated disc and the source on the opposite side of the disc, partially visible through the optically-thin, central region.

We also demonstrated how the LP models fitted with a free normalization of the reflected component are affected when the model with a self-consistent description of the non-isotropy of the X-ray source is applied to explain the reduction of reflection strength. The related effects are most apparent in models with low $h$, where strong beaming is needed to counteract the light bending, e.g.\ in the model for Cyg X-1 of \citetalias{2015ApJ...808....9P}. This model requires $\beta > 0.8$, significantly altering the relativistic blurring, which can be only partially compensated by the change of the model parameters and gives a poor fit formally disfavoring the {\tt eqpair} solution\footnote{Another argument against this solution, namely, the unrealistic size of the X-ray source implied by the fitted {\tt eqpair} parameters, was pointed out by \citetalias{2017MNRAS.472.4220B}.} with an outflow.
Similarly, the self-consistent inclusion of the non-isotropy instead of the free scaling of reflection in the model with stratified Comptonization gives significantly different parameters, 
e.g.\ $\xi$ and $Z_{\rm Fe}$ lower by over a factor of 2, and much larger $i$ compared to those reported by \citetalias{2017MNRAS.472.4220B}.

We conclude that results obtained using the LP model with a free scaling of reflection should be treated with extreme care and considered as nonphysical (i.e.\ purely phenomenological) unless a self-consistent explanation for the reduction of reflection strength is given. Currently, such an explanation is lacking. This reduction can be self-consistently attributed to a non-isotropic emission only if the related change of the spectral shape is taken into account. We directly demonstrated this for the kinematic beaming, but the same is obviously true for any mechanism leading to a non-isotropic emission.
Alternatively, the observed reflection could be lessened due to the Comptonization of reflection in the X-ray source \citep[e.g.][]{2017ApJ...836..119S}. This, however, would require the X-ray corona covering a large part of the inner disc, at variance with the assumptions of the LP model.
This model is typically applied assuming a point-like source, representing a very compact corona.
Indeed, when models allowing to fit the corona size are applied, e.g.\ {\tt reflkerrV} presented here or {\tt reflkerr\_elp} of \citet{2020A&A...641A..89S},
it is found that the corona must be very compact, with the size of the order of a gravitational radius for the model of \citetalias{2015ApJ...808....9P} for Cyg X-1 and similarly for the Seyfert galaxy 1H0707-495 \citep[see][similar constraints for other objects will be presented in our future work, Klepczarek et al., in preparation]{2020A&A...641A..89S}, to reproduce the results obtained with a point-like LP. Such compact coronae intercept a small fraction of reflected photons, making their Comptonization a negligible effect.

\section*{Acknowledgements}
We thank A.\ Zdziarski for comments. This research has been supported in part by the Polish National Science Centre grants 2016/21/B/ST9/02388 
and 2019/35/B/ST9/03944.

\section*{Data Availability}
The data are publicly available at NASA's HEASARC. {\tt reflkerrV} is available at \url{https://wfis.uni.lodz.pl/reflkerr}

\bibliographystyle{mnras}
\bibliography{lpV2}{}

\label{lastpage}
\end{document}